\documentclass[aps,pra,twocolumn,amsmath,amssymb,nofootinbib,superscriptaddress]{revtex4-1}

\newcommand{\expect}[1]{\left\langle{#1}\right\rangle}

\newcommand{\set}[1]{\left\lbrace #1 \right\rbrace}

\usepackage[colorlinks,linkcolor=blue,citecolor=blue,allcolors=blue]{hyperref} 
\usepackage[pdftex]{graphicx}
\usepackage{subfigure}
\usepackage{mathrsfs}
\usepackage{subfigure}
\usepackage{float}
\usepackage[usenames,dvipsnames]{xcolor}
\usepackage{framed}

\definecolor{shadecolor}{rgb}{1.0, 0.99, 0.82}

\begin{document}

\title{Logarithmic Subdiffusion from a Damped Bath Model}

\author{Thomas Guff}
\email[]{t.guff@surrey.ac.uk}
\affiliation{School of Mathematics and Physics, University of Surrey, GU2 7XH Guildford, United Kingdom}

\author{Andrea Rocco}
\email[]{a.rocco@surrey.ac.uk}
\affiliation{School of Mathematics and Physics, University of Surrey, GU2 7XH Guildford, United Kingdom}
\affiliation{School of Biosciences, University of Surrey, GU2 7XH Guildford, United Kingdom}

\date{\today}

\begin{abstract}

A damped oscillator heat bath model is a modification of the standard heat bath model, wherein each bath oscillator itself has a Markovian coupling to its own heat bath \cite{Plyukhin2019}. We modify such a model to one where the resulting damping of the oscillators is linear in their frequency rather than being a constant. We find that this generates a memory kernel which behaves like $k(t) \sim 1/t$ as $t\to \infty$, which is a boundary case not considered in previous works.
As the memory kernel does not have a finite integral, the reduced system is subdiffusive, and we numerically show that diffusion goes as $\expect{\Delta Q^{2}(t)} \sim t/\log(t)$ as $t\to \infty$. We also numerically calculate the velocity correlation function in the asymptotic regime and use it to confirm the aforementioned subdiffusion.

\end{abstract}

\maketitle

\section{Introduction}\label{sec:intro}

Brownian motion gives rise to so-called normal diffusion, where the mean-squared displacement $\expect{\Delta Q^{2}} = \expect{(Q-\expect{Q})^{2}}$ of the position $Q$ of a free particle is asymptotically linear in time.
This motion is found in nature: for example in the random movement of a large particle, such as pollen, suspended in a liquid or gaseous medium, as first modelled by Einstein \cite{Einstein1905}.

However, some systems will exhibit superdiffusion, where the mean-squared displacement increases faster than linearly in the asymptotic time regime. Interesting examples include atmospheric eddies such as smoke and turbulence \cite{Richardson1926}.
Conversely, other systems exhibit subdiffusion, where the mean-squared displacement is sub-linear in time. This can be due to crowding effects such as those occurring in biological systems \cite{Weiss2004,Regner2013}. These are all examples of anomalous diffusion and are typically described in terms of a power law mean-square displacement $\expect{\Delta Q^{2}(t)} \sim t^{\alpha}$, where $\alpha>1$ corresponds to superdiffusion and $\alpha < 1$ indicates subdiffusion.

Diffusion is studied in both classical and quantum systems. In both cases normal diffusion is effectively modelled using the classical or quantum Langevin equation driven by a thermal stochastic force.

Along the same lines, anomalous diffusion can also be modelled classically by modifying the Langevin equation to achieve non-standard properties. One possibility is to modify the statistics of the stochastic force \cite{Jespersen1999}. Alternatively, one can replace the dissipation term with a convolution integral of the velocity and a memory kernel --- leading to the so-called generalised Langevin equation --- and then study the diffusive effects induced by different memory kernels \cite{Porra1996}. In particular, if the memory kernel behaves asymptotically as $k(t) \sim t^{-\alpha}$ where $0 < \alpha < 1$, then the system will be  subdiffusive, with $\expect{\Delta Q^{2}(t)} \sim t^\alpha$ \cite{Plyukhin2019,Kupferman2004}.

The Langevin equation can also be modified by replacing ordinary derivatives with fractional derivatives \cite{Grigolini1999,Lutz2001}, and the same approach can be used 
for the Fokker-Planck equation \cite{Metzler1999,Metzler2000}.

Similarly, in the quantum case one can introduce fractional derivatives in the master equation to study anomalous quantum diffusion \cite{Tarasov2013}. However, quantum mechanical systems are more subtle due to low temperature quantum effects, which often change the diffusion properties; indeed a high-temperature limit is often taken for simplicity.

Both classically and quantum mechanically, the generalised Langevin equation can be derived from an open systems model where a system of interest interacts with a bath composed of harmonic oscillators \cite{Ford1965,Cortes1985,Caldeira1983}.
Anomalous diffusion can then be modelled by appropriately changing the interaction between system and bath, that is the distribution of oscillator frequencies and the coupling between each oscillator and the system.
In particular, the choice of an Ohmic spectral density, linear in frequency with an infinite frequency cut-off, leads to delta correlated noise and Markovian evolution.
More generally, different power-law spectral densities correspond to more exotic diffusive behaviour, and the corresponding mean-squared displacement does not necessarily follow a power-law \cite{Grabert1987}. Fractional coupling between the bath and system \cite{Vertessen2023}, random interaction Hamiltonians \cite{Lutz2001,Lutz2001b}, or a zero-temperature heat bath in quantum mechanical systems \cite{BreuerPetruccione07} also produce anomalous features.

\begin{figure}[b!]
\includegraphics[scale=0.5]{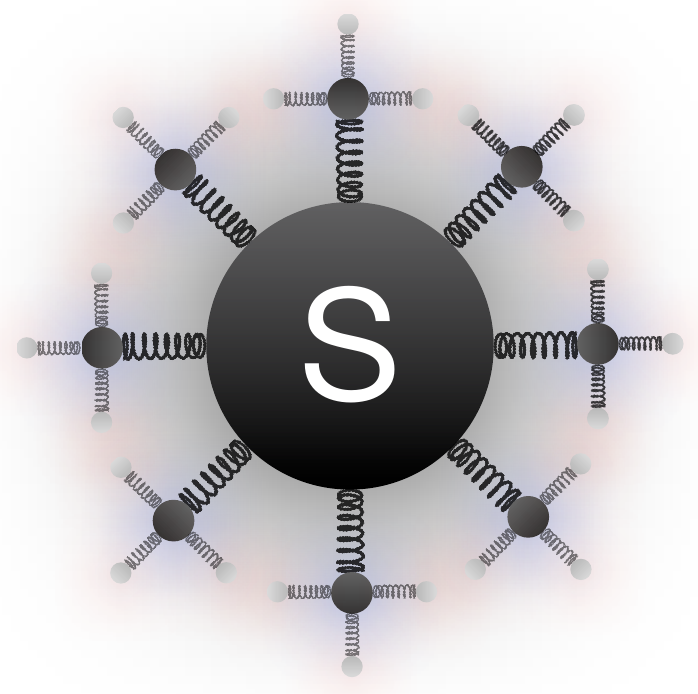}
\caption{Schematic of the damped bath model. The system of interest is coupled to a collection of harmonic oscillators, each of which is in turn coupled to their own set of harmonic oscillators. The secondary baths induce Markovian dissipation in each oscillator of the primary bath. This modified bath structure leads to subdiffusion without changing the nature of coupling (i.e. the spectral density) between the system of interest and the primary bath.}
\label{fig:schematic}
\end{figure}

An alternative approach is proposed in \cite{Plyukhin2019}, whereby  the internal structure of the bath is changed: using instead a damped bath model to generate a memory kernel with the aforementioned power law.
This model consists of a system coupled to a bath of harmonic oscillators, with each oscillator in turn coupled to its own bath of harmonic oscillators (see fig.~\ref{fig:schematic}).
The interaction between each primary bath oscillator and it's corresponding bath is presumed to be Markovian. One can consider this damped bath model as a particle interacting with a bath of \emph{Brownian particles} instead of harmonic oscillators. In \cite{Plyukhin2019}, each primary oscillator undergoes the same damping. 

In this paper we show that a simple modification to this model can be used to generate a subdiffusive boundary case. We modify the model to allow the damping of each primary bath oscillator to be proportional to the frequency.
With this modification the memory kernel behaves asymptotically as $k(t)\sim 1/t$ even with a linear (Ohmic) spectral density; a non-linear power-law spectral density is not required.
It is the internal structure of the bath which causes it to respond to the system according to an infinite time-scale, leading to anomalous diffusion.

This power-law behaviour is a boundary case  of the range considered above, and is not readily represented as a fractional derivative.
However since it is non-integrable, it still results in subdiffusion \cite{Morgado2002}, although not governed by a power-law.
We show numerically that the mean-squared displacement asymptotically behaves like $\expect{\Delta Q^{2}(t)} \sim t/\log(t)$, and confirm this diffusion using the velocity autocorrelation function.
This diffusive behaviour does not arise from the standard Caldeira-Leggett model with a power-law spectral density, at any temperature \cite{Grabert1987}.

\section{Damped Bath Model} \label{sec:damped}
\subsection{Memory kernel}
The model consists of a single position degree of freedom $Q$, and associated momentum $P$, coupled linearly to a bath of $N$ harmonic oscillators $\set{q_{k},p_{k}}_{k=1}^{N}$.
The Hamiltonian consists of three terms
\begin{equation}
H = H_{\textsc{s}}+H_{\textsc{b}} + H_{\textsc{int}}.
\end{equation}
The system and interaction parts are given by
\begin{subequations}
\begin{align}
H_{\textsc{s}} &= \frac{P^{2}}{2M} + \sum_{n=1}^{N}\frac{g_{n}^{2}}{2 m_{n}\omega_{n}^{2}}Q^{2} \\
H_{\textsc{int}} &= -\sum_{n=1}^{N} g_{n} q_{n} Q,
\end{align}
\end{subequations}
where the quadratic system term counters the potential re-normalisation induced from the interaction of the system and the bath.
The bath Hamiltonian term $H_{\textsc{b}}$ has the usual kinetic and quadratic potential terms but to each oscillator we add an extra Hamiltonian term $H_{\textsc{b}}^{(n)}$
\begin{equation}
H_{\textsc{b}} = \sum_{n=1}^{N} \left(\frac{p_{n}^{2}}{2m_{n}} + \frac{m_{n} \omega_{n}^{2} q_{n}^{2}}{2} + H_{\textsc{b}}^{(n)}\right),
\end{equation}
where $H_{\textsc{b}}^{(n)}$ is an entire secondary harmonic bath Hamiltonian consisting of $N_{n}$ oscillators coupled linearly to the $n^{\text{th}}$ oscillator,
\begin{equation}
H_{\textsc{b}}^{(n)} = \sum_{\nu=1}^{N_{n}} \left(\frac{p_{n,\nu}^{2}}{2m_{n,\nu}} + \frac{m_{n,\nu}\omega_{n,\nu}^{2}}{2}\left(q_{n,\nu} - \frac{g_{n,\nu}}{m_{n,\nu}\omega_{n,\nu}^{2}}q_{n}\right)^{2}\right).
\end{equation}
Here $q_{n,\nu}$ and $p_{n,\nu}$ are the position and momentum respectively of the $\nu^{\text{th}}$ oscillator in the secondary bath coupled to the $n^{\text{th}}$ oscillator of the primary bath.
Instead of a standard harmonic equation of motion, the equation of motion of the $n^{\text{th}}$ primary bath oscillator takes the form of a generalised Langevin equation, owing to the interaction with the secondary bath,
\begin{align}
\ddot{q}_{n} + \omega_{n}^{2}q_{n}+ \int_{0}^{t} k_{n}(t-t^{\prime})&\dot{q}_{n}(t^{\prime})\, dt^{\prime} + k_{n}(t)q_{n}(0) \nonumber\\
&=\frac{1}{m_{n}}\varepsilon_{n}(t) + \frac{g_{n}}{m_{n}}Q.\label{eq:GLEb}
\end{align}
The memory kernel $k_{n}$ and the stochastic force $\varepsilon_{n}$ are given by the familiar expressions for the standard oscillator environmental model \cite{BreuerPetruccione07}.
The stochasticity of $\varepsilon_{n}(t)$ arises from the (thermal) uncertainty of the initial conditions $q_{n,\nu}(0)$ and $p_{n,\nu}(0)$.

Assuming a standard Ohmic spectral density for each secondary bath, the  memory kernels become proportional to the Dirac delta function,
\begin{equation}
    k_{n}(t) = 2\gamma_{n} \delta(t).\label{eq:newr}
\end{equation}
The damping coefficients $\gamma_{n}$ here are free parameters and correspond to the choice of slope near zero of the spectral density of each secondary bath.
After inserting \eqref{eq:newr} into \eqref{eq:GLEb}, the primary bath equation of motion becomes that of a damped and driven harmonic oscillator,
\begin{equation}
\ddot{q}_{n} + 2\gamma_{n} \dot{q_{n}} +\omega_{n}^{2}q_{n} = \frac{g_{n}}{m_{n}}Q + \frac{1}{m_{n}}\varepsilon_{n}(t). \label{eq:inhomogeneous}
\end{equation}

Since the probability distribution for the initial conditions is thermal at temperature $T$, the stochastic forces $\varepsilon_{n}(t)$ have the usual white noise statistics,
\begin{equation}
\expect{\varepsilon_{n}(t)} = 0, \; \; \expect{\varepsilon_{n}(t)\varepsilon_{j}(t^{\prime})} = 4k_{B} T m_{n} \gamma_{n}\delta(t-t^{\prime})\delta_{ij} \label{eq:2ndstat}
\end{equation}
where $k_{B}$ is Boltzmann's constant. 
The homogeneous part of the solution to \eqref{eq:inhomogeneous} is
\begin{equation}
q_{n}^{(h)}(t) = a_{n}(t)q_{n}(0) + b_{n}(t)p_{n}(0),
\end{equation}
where
\begin{subequations}
\begin{align}
a_{n}(t) &= e^{-\gamma_{n} t}\left(\cos\Omega_{n} t + \frac{\gamma_{n}}{\Omega_{n}}\sin\Omega_{n}t\right),\\
b_{n}(t) &= \frac{1}{m_{n}\Omega_{n}}e^{-\gamma_{n} t}\sin\Omega_{n}t,
\end{align}
and $\Omega_{n}^{2} = \omega_{n}^{2} - \gamma^{2}_{n}$.
The Green's function for the particular solution is $\sum_{n} g_{n} b_{n}(t)$; and so the equation of motion for the system of interest is itself a generalised Langevin equation,
\end{subequations}
\begin{align}
M\frac{d^{2}Q}{dt} + \int_{0}^{t} &k(t-t^{\prime})\dot{Q}(t^{\prime}) \, dt^{\prime}+ k(t)Q(0) = f(t) \label{eq:Lgvn}
\end{align}
with a memory kernel and stochastic force which show the effect of the secondary bath coupling,
\begin{subequations}
\begin{align}
k(t) &= \sum_{n=1}^{N} \frac{g_{n}^{2}}{m_{n}\omega_{n}^{2}}a_{n}(t), \label{eq:k(t)}\\
f(t) &= \sum_{n=1}^{N} \left(g_{n}q_{n}^{(h)}(t) + g_{n}\int_{0}^{t}b_{n}(t-t^{\prime})\varepsilon_{n}(t^{\prime})\,  dt^{\prime}\right).
\end{align}
\end{subequations}
\begin{figure}[t!]
\includegraphics[scale=0.6]{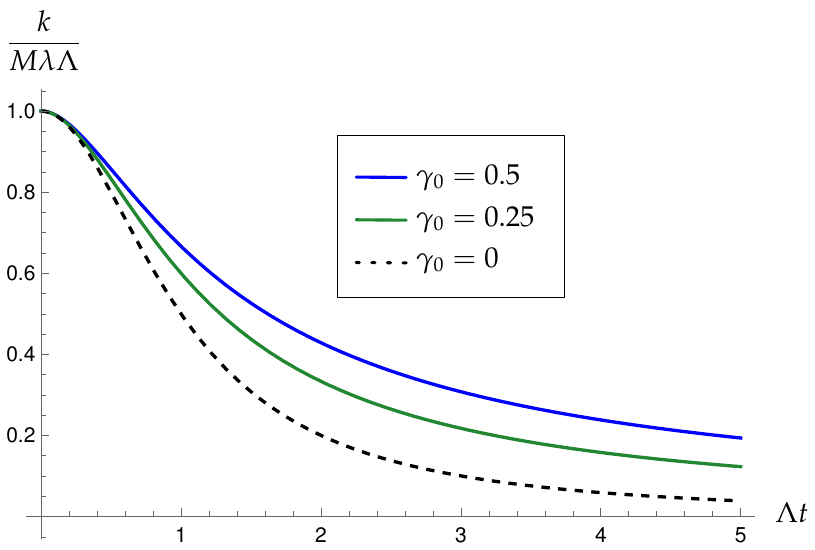}
\caption{Memory kernel for the damped bath model with an Ohmic spectral density with an exponential cut-off. The kernel is plotted for $\gamma_{0}=0,0.25,0.5$. When $\gamma_{0}=0$, then $k(t)\sim 1/t^{2}$, and the kernel has a finite integral over the positive real line, giving a finite bath time-scale. When $\gamma_{0}>0$, then $k(t)\sim 1/t$, which has an infinite time scale, leading to subdiffusion.}
\label{fig:kernel}
\end{figure}
Assuming the initial condition of the primary bath is a thermal state at temperature $T$ (the same temperature as the secondary baths) in the quantum case, or has thermally distributed initial conditions, the fluctuation-dissipation relation holds,
\begin{equation}
\frac{1}{2}\expect{\set{f(t),f(t^{\prime})}} = k_{B} T k(|t-t^{\prime}|),\label{eq:fdt}
\end{equation}
where $\set{\cdot,\cdot}$ denotes the anti-commutator.
In the quantum case the high-temperature limit must be taken so that the ground state fluctuations become negligible.

We now take the usual continuum limit of the primary bath.
To do this, we re-write the memory kernel in the integral form,
\begin{equation}
k(t) = \int_{0}^{\infty} \frac{J(\alpha)}{\omega(\alpha)} a_{\alpha}(t)\, d\alpha \label{eq:kcont}
\end{equation}
where $\alpha$ is a dimensionless continuous index with range $[0,\infty)$ and $a_{\alpha}(t)$ is simply $a_{n}(t)$ with $\gamma_{n}\to \gamma(\alpha)$, and $\Omega_{n} \to \Omega(\alpha)$; and $J(\alpha)$ is the spectral density
\begin{equation}
J(\alpha) = \sum_{n=1}^{N} \frac{g(\alpha)^{2}}{m(\alpha)\omega(\alpha)} \delta(\alpha - n), \label{eq:fullspec}
\end{equation}
where $m(\alpha)$ and $g(\alpha)$ are the continuous versions of $m_{n}$ and $g_{n}$ respectively; and $\delta(\alpha)$ is the Dirac delta function.
The spectral density \eqref{eq:fullspec}, if inserted into \eqref{eq:kcont} will yield the original discrete memory kernel \eqref{eq:k(t)}.

The Ohmic approximation is now done with the following steps.
First it is standard to assume the mass is constant across all oscillators, and the frequency grows linearly in the index $\alpha$,
\begin{equation}
    m(\alpha) = m, \qquad \omega(\alpha) = \omega_{0}\alpha,
\end{equation}
where $\omega_{0}$ is a constant with units of frequency.
We now assume that the spectral density is linear in the frequency $\omega(\alpha)$, with exponential high-frequency cut-off,
\begin{equation}
    J(\alpha) = m \lambda_{0}^{2} \omega_{0} \alpha e^{-\frac{\omega_{0}}{\Lambda}\alpha},
\end{equation}
where $\Lambda = \omega_{0} N$ is the frequency of the oscillator with index $N$, and $\lambda_{0}$ is a dissipation constant with units of frequency.

As mentioned above, the damping rate $\gamma(\alpha)$ varies with each primary oscillator. This variation corresponds to different choices of the slope of the Ohmic spectral density for each secondary bath.
In contrast to \cite{Plyukhin2019}, we choose a damping rate linear in $\alpha$:
\begin{equation}
    \gamma(\alpha) = \omega_{0}\gamma_{0} \alpha. \label{eq:lineardamping}
\end{equation}
Here $\gamma_{0}$ is a dimensionless constant which we have inserted to simplify later equations, but can be thought of as the ratio
\begin{equation}
    \gamma_{0} = \frac{\gamma(\alpha)}{\omega(\alpha)},
\end{equation}
which is a constant since both $\gamma(\alpha)$ and $\omega(\alpha)$ are linear in $\alpha$.
We note that $\gamma_{0} = 0$ corresponds to no coupling between the primary and secondary baths, in which case the model reverts back to the standard Caldeira-Leggett model.
For convenience we assume $0 \leq \gamma_{0} < 1$ but this is not a necessary restriction.


With these assumptions, the shifted frequencies $\Omega(\alpha)$ are given by,
\begin{equation}
    \Omega(\alpha) = \sqrt{\omega_{0}^{2}\alpha^{2} - \gamma_{0}^{2}\omega_{0}^{2}\alpha^{2}} = \gamma_{1}\omega_{0}\alpha,
\end{equation}
where $\gamma_{1} = \sqrt{1-\gamma_{0}^{2}}\,$; and $a_{\alpha}(t)$ can be written
\begin{equation}
    a_{\alpha}(t) = e^{-\gamma_{0}\omega_{0}\alpha t}\left( \cos(\gamma_{1}\omega_{0}\alpha) + \frac{\gamma_{0}}{\gamma_{1}} \sin(\gamma_{1}\omega_{0}t)\right).
\end{equation}

The memory kernel can now be calculated,
\begin{align}
    k(t) &= M\lambda_{0}^{2} \int_{0}^{\infty} e^{-\frac{\omega_{0}}{\Lambda}\alpha} a_{\alpha}(t) \, d\alpha \nonumber \\
    &= \frac{M\lambda \Lambda (1+2\gamma_{0}\Lambda t)}{1+2\gamma_{0}\Lambda t + \Lambda^{2}t^{2}}, \label{eq:kernel}
\end{align}
where we have introduced $\lambda = \lambda_{0}^{2}/\omega_{0}$ for simplicity.
Asymptotically, this memory kernel behaves as $k(t) \sim 1/t$ as $t \to \infty$.
If we choose $\gamma_{0}=0$ (that is turn off the interaction between the primary and secondary baths), then the kernel reverts to $k(t) \sim 1/t^{2}$, which has a finite integral on $[0,\infty)$. 
It is therefore the presence of the secondary bath which causes the time-scale in which the bath responds to the system of interest to be infinite (see fig.~\ref{fig:kernel}).

\subsection{Effect of the secondary bath on the possible types of diffusion}

The type of diffusion arising from a given memory kernel can be determined from the integral of that memory kernel over the positive real axis \cite{Morgado2002}.
In our case we integrate \eqref{eq:k(t)}
\begin{align}
    \int_{0}^{\infty} k(t) \, dt &= \sum_{n=1}^{N} \frac{2\gamma_{n} g^{2}_{n}}{m_{n} \omega_{n}^{4}} \nonumber \\
    &= \int_{0}^{\infty} \frac{2\gamma(\alpha)J(\alpha)}{\omega(\alpha)^{3}} \, d\alpha, \label{eq:DiffType}
\end{align}
where we used \eqref{eq:fullspec} to write the result in integral form.
The diffusion will be subdiffusive, standard, or superdiffusive when \eqref{eq:DiffType} is infinite, positive, or zero respectively. \cite{Morgado2002}.
If we assume that the spectral density contains a high-frequency cut-off such that the tail of $\gamma(\alpha)J(\alpha)/\omega(\alpha)^{3}$ is always integrable (such as an exponential cut-off), then the diffusion is determined by the behaviour near $\alpha = 0$.
If we write
\begin{equation}
    \frac{\gamma(\alpha)J(\alpha)}{\omega(\alpha)^{3}} \sim \alpha^{n}, \quad \text{as } \alpha\to 0,
\end{equation}
then subdiffusion occurs for $n \leq -1$, and standard diffusion occurs for $n > -1$.
Our choice of $\gamma(\alpha)$, $\omega(\alpha)$ and $J(\alpha)$ to be linear in $\alpha$ as $\alpha \to 0$ results in $\gamma(\alpha)J(\alpha)/\omega(\alpha)^{3} \sim \alpha^{-1}$ as $\alpha \to 0$, implying subdiffusion.

Note that since $J(\alpha), \omega(\alpha), \gamma(\alpha) > 0$, there is no choice of spectral density which could make \eqref{eq:DiffType} zero (even using a minimal frequency cut-off as suggested in \cite{Morgado2002} for the standard Caldeira-Leggett model).
That is, there is no way to generate superdiffusion using a damped bath model.

\section{Subdiffusion}
\begin{figure}[t!]
\subfigure[]{\includegraphics[scale=0.6]{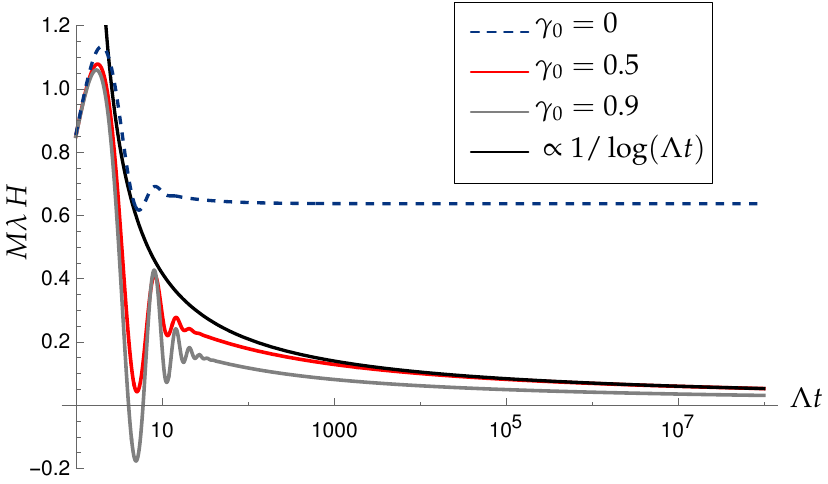}}\\
\subfigure[]{\includegraphics[scale=0.6]{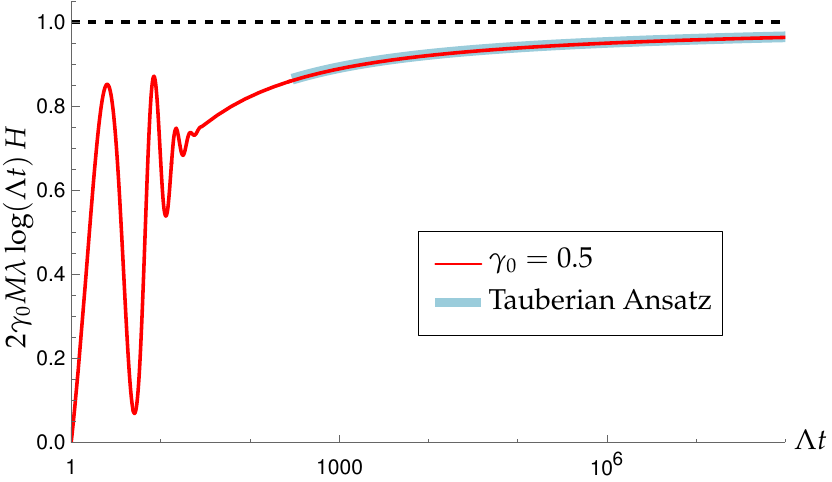}}
\caption{(a) The Green's function $H$ \eqref{eq:H} for the memory kernel \eqref{eq:kernel}, computed numerically for values $\gamma_{0}=0,0.5,0.9$, along with the asymptotic fit of $1/\log(t)$ (black) which should apply to the $\gamma_{0}=0.5$ case. When $\gamma_{0}=0$, the secondary bath is turned off and the Green's function approaches a constant value, as in standard theory. In the asymptotic regime the fit is very accurate. (b) The Green's function normalised by the inverse logarithm function (red) showing convergence to unity. In blue is plotted the inverse Laplace transform of \eqref{eq:Tauberian}, showing agreement in the asymptotic limit.}
\label{fig:H}
\end{figure}

The generalised Langevin equation \eqref{eq:Lgvn} can be solved using Laplace transforms \cite{Porra1996,Ford2001}, and from this, the mean-squared displacement $\expect{Q^{2}(t)}$ is given by
\begin{align}
\expect{\Delta Q^{2}(t)} &= \int_{0}^{t}\int_{0}^{t} H(t-t_{1})H(t-t_{2}) \nonumber \\ 
&\qquad\qquad\times\expect{\set{f(t_{1}),f(t_{2})}} \, dt_{1}dt_{2}  \label{eq:Q2}\\
&= 2 k_{B}T	\int_{0}^{t}\int_{0}^{t} H(t_{1})H(t_{2})k(|t_{1}-t_{2}|) \, dt_{1}dt_{2}. \nonumber 
\end{align} 
Here the Green's function $H(t)$ is defined through its Laplace transform,
\begin{equation}
\int_{0}^{\infty} H(t) e^{-st} \, dt = \frac{1}{Ms^{2}+s\tilde{k}(s)}, \label{eq:H}
\end{equation}
where $\tilde{k}(s)$ is the Laplace transform of the memory kernel.
In the case of \eqref{eq:kernel}, we have the expression
\begin{align}
    \tilde{k}(s) =  -\frac{M\lambda}{\gamma_{1}} &e^{\frac{s}{\Lambda}(\gamma_{0}-i \gamma_{1})} (\gamma_{0}-i\gamma_{1})^{2} \\
    &\times\left(i \, \text{E}_{1}\left(\frac{s}{\Lambda}(\gamma_{0}-i\gamma_{1})\right)\right) +  \textsc{c.c.},\nonumber
\end{align}
where $\text{E}_{1}$ denotes the exponential integral and $\textsc{c.c.}$ denotes the complex conjugate. Results from Tauberian theory \cite{Feller1991} connect the asymptotic behaviour of a function to the behaviour of its Laplace derivative around zero.
In this spirit we see that $\tilde{k}(s)$ behaves as $\tilde{k}(s) \sim -2M\lambda \gamma_{0} \log(s)$ as $s\to 0$.
Consequently,
\begin{equation}
\frac{1}{Ms^{2}+s \tilde{k}(s)} \sim -\frac{1}{2M\lambda \gamma_{0} s\log(s)} \quad \text{as} \;  s \to 0. \label{eq:Tauberian}
\end{equation}
\begin{figure}[b!]
\includegraphics[scale=0.6]{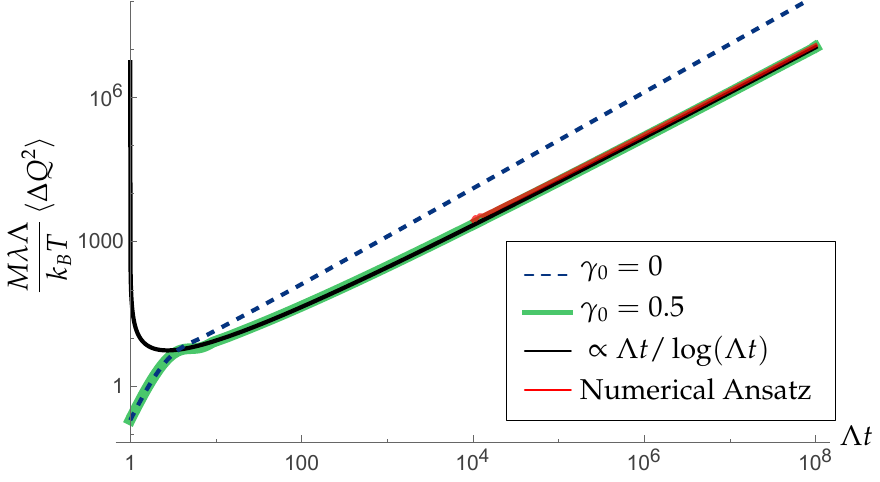}
\caption{Log-log plot of the mean-squared displacement $\expect{\Delta Q^{2}}$ calculated numerically using \eqref{eq:Q2} where $H$ was computed numerically as in fig.~\ref{fig:H} with a choice of $\gamma_{0}=0.5$ (green) and $\gamma_{0}=0$ (blue, dashed). When $\gamma_{0}=0$, the system undergoes standard linear diffusion. When $\gamma_{0}>0$ the system undergoes subdiffusion, asymptotically as $\expect{\Delta Q^{2}} \sim t/\log(t)$. This behaviour is fitted in black. In red is the result of choosing $H$ to be precisely the asymptotic expression \eqref{eq:Hparam}; which has a highly oscillatory initial phase but which gives the same asymptotic subdiffusion.}
\label{fig:Q2}
\end{figure}

The Green's function $H(t)$ \eqref{eq:H} cannot be calculated analytically but can be evaluated numerically. 
In fig.~\ref{fig:H} we have plotted $H(t)$ corresponding to the memory kernel $k(t)$ \eqref{eq:kernel} in the asymptotic limit.
We can see by plotting a fitting curve that asymptotically it behaves as $H(t) \sim 1/\log(t)$ as $t\to 0$.
The same asymptotic behaviour is seen when numerically calculating the inverse Laplace transform of \eqref{eq:Tauberian}.
The form of \eqref{eq:Tauberian} suggests that the the parameter dependence is
\begin{equation}
H(t) \sim \frac{1}{2M \lambda \gamma_{0} \log(\Lambda t)},\label{eq:Hparam}
\end{equation}
and this is confirmed numerically.
When $\gamma_{0}=0$, the memory kernel becomes integrable, and $H(t)$ asymptotes to a constant value $2/M \lambda \pi$.

\begin{figure}[t!]
\includegraphics[scale=0.55]{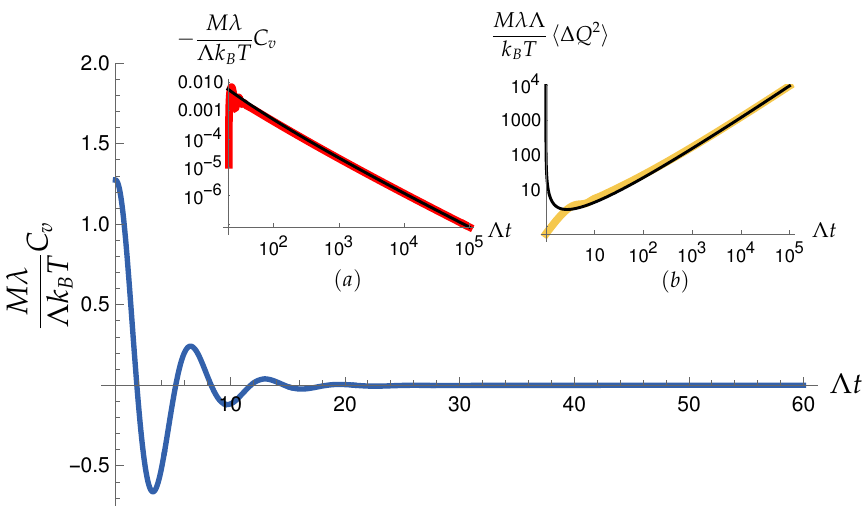}
\caption{Velocity correlation function $C_{v}$ on linear scales as a function of the dimensionless time parameter $\Lambda t$. Numerically calculated using \eqref{eq:v}, with $\dot{H}$ derived from $H$, numerically computed in fig.~\ref{fig:H}. Here $\gamma_{0}=0.5$, and $\Lambda t_{0}$ was chosen to be large enough such that $C_{v}$ became independent of $\Lambda t_{0}$ (in these plots it was set to $\Lambda t_{0}=10^{5}$. Insets: (a) log-log plot of $-C_{v}$, (red) and a fit of $1/(\Lambda t  \log(\Lambda t)^{2})$ (black) showing the asymptotic behaviour of $C_{v}$. (b) Log-log plot of the variance of position $\expect{\Delta Q^{2}}$ numerically integrated using \eqref{eq:vdiff} with $C_{v}$ itself numerically calculated as above (yellow) and a fit of $\Lambda t/\log(\Lambda t)$ (black). This confirms the logarithmic subdiffusive behaviour.}
\label{fig:Cv}
\end{figure}
These numerically determined curves can then be substituted into \eqref{eq:Q2} and then a numerical integration can be used to determine $\expect{\Delta Q^{2}}$.
The results are plotted in fig.~\ref{fig:Q2}: when there is no secondary bath ($\gamma_{0}=0$) the diffusion has the standard linear form.
The presence of the secondary bath causes logarithmic subdiffusion, with asymptotic behaviour,
\begin{equation}
\expect{\Delta Q^{2}} \sim \frac{k_{B} T}{2\gamma_{0}^{2} M \lambda \Lambda}\frac{\Lambda t}{\log(\Lambda t)}, \label{eq:result}
\end{equation}
as $t\to \infty$.
This is in numerical agreement with the resulting diffusion that is achieved by simply assuming that $H$ takes its asymptotic form \eqref{eq:Hparam} for all time.

To confirm this asymptotic behaviour, we examine the velocity correlation function,
\begin{equation}
C_{v}(t) = \expect{v(t_{0}+t)v(t_{0})}. \label{eq:Cv}
\end{equation}
Here it is assumed that the system is in a sufficiently asymptotic regime that $C_{v}$ is stationary, that is the right hand side of \eqref{eq:Cv} is independent of $t_{0}$.
In such a regime there is a well-known relationship between $C_{v}$ and the mean-squared displacement \cite{Kubo1966,Muralidhar1990,Morgado2002},
\begin{equation}
\expect{\Delta Q^{2}(t)} = \int_{0}^{t} (t-s) C_{v}(s) \, ds. \label{eq:vdiff}
\end{equation}
The velocity of the system is given by
\begin{equation}
v(t) = \ddot{H}(t)x(0) + \dot{H}(t)v(0) + \int_{0}^{t} \dot{H}(t-t^{\prime})f(t^{\prime}). \label{eq:v}
\end{equation} 
In calculating the correlation function, the homogeneous terms of \eqref{eq:v} will be negligible when $t_{0}$ is large, since they're all $o(1/t)$.
Likewise the cross terms will average to zero, on the assumption that the system initial conditions are uncorrelated with $f(t)$.
Therefore only the integral terms are relevant
\begin{align}
&\langle v(t_{0}+t)v(t_{0})\rangle \\
&= 2k_{B}T\int_{0}^{t_{0}}\int_{0}^{t_{0}+t} \dot{H}(t^{\prime})\dot{H}(t^{\prime\prime}) k(|t+t^{\prime}-t^{\prime\prime}|) dt^{\prime\prime}  dt^{\prime}, \nonumber
\end{align}
where we have used \eqref{eq:fdt}.
This integral can be computed numerically, and this is plotted in fig.~\ref{fig:Cv}.
We see an initial oscillatory period which relaxes to a negative decaying function. Through numerical fitting we can see that the asymptotic behaviour is
\begin{equation}
C_{v}(t) \sim \frac{\Lambda k_{B} T}{2 \gamma_{0}^{2} M\lambda}\frac{-1}{\Lambda t (\log(\Lambda t))^{2}} \quad \text{as } t\to\infty.\label{eq:Casym}
\end{equation}
This asymptotic behaviour means that $C_{v}(t)$ has a finite integral over the non-negative reals. However it approaches zero slower than $o(t^{-2})$, and hence it does not meet the criteria for standard diffusion \citep{Muralidhar1990}.
We can then use \eqref{eq:vdiff} to compute $\expect{\Delta Q^{2}}$ to show the same subdiffusive behaviour as in \eqref{eq:result} (see fig.~\ref{fig:Cv}).
Indeed, if we simply use the asymptotic form of $C_{v}$ \eqref{eq:Casym} in \eqref{eq:vdiff}, beginning the integration at some time $t_{0} \gg 1$, then we have
\begin{equation}
\int_{t_{0}}^{t}(t-\tau)C_{v}(\tau) d\tau = \frac{k_{B}T}{2\gamma_{0}^{2} M\lambda \Lambda} \text{li}(\Lambda t) - h(t,t_{0}),
\end{equation}
where $\text{li}(\Lambda t)$ is the logarithmic integral, and $h(t,t_{0})$ is a linear function in $t$. The linear function $h$ is cancelled by the contribution of the oscillating part of $C_{v}$ on $[0,t_{0}]$, and asymptotically, we have
\begin{equation}
\frac{k_{B}T}{2\gamma_{0}^{2}M \lambda \Lambda} \text{li}(t) \sim \frac{k_{B}T}{2\gamma_{0}^{2}M \lambda \Lambda} \frac{\Lambda t}{\log(\Lambda t)} \quad \text{as }t\to\infty.
\end{equation}
This provides a further confirmation of the logarithmic subdiffusion that we found numerically above.

\section{Conclusion}

The damped bath structure provides a Hamiltonian model which exhibits anomalous diffusion. 
Rather than having unusual coupling between system and environment, it simply modifies the internal structure of the bath itself, which causes the memory effects which lead to subdiffusion.
Indeed, even though the bath is in equilibrium, the internal bath structure allows for an infinite time scale that characterizes the dynamics.

We saw that a minor modification of the structure described in \cite{Plyukhin2019} caused the system to have a fairly exotic subdiffusive behaviour.
In particular we assumed the dissipation constant for each bath oscillator to be proportional to its frequency, with all oscillators having the same damping proportionality constant $\gamma_{0}$. 
This led to a memory kernel with asymptotic behaviour $k(t)\sim 1/t$, even with the typical Ohmic spectral density.
While the very high frequency oscillators would thus have to dissipate extremely quickly, these oscillators have very small effect on the system of interest due to the exponential cut-off.
We note here that it is insignificant whether the oscillators are underdamped or overdamped (or even perfectly damped), since all of these cases result in the same asymptotic memory kernel.

Memory kernels often considered in the literature have the asymptotic form $t^{-\alpha}$ where $0 < \alpha < 1$, as this results in subdiffusion which itself follows a power-law.
Our memory kernel is thus an interesting boundary case which does not result in power-law diffusion.
It is also at the boundary between memory kernels which have a finite and infinite time-scale, the former resulting in standard diffusion.
Therefore the diffusion in our damped bath model of $\expect{\Delta Q^{2}(t)}\sim t/\log(t)$ may be described as the fastest possible subdiffusion.
An interesting question is whether this diffusive behaviour can be replicated with the standard bath-oscillator model with a non-standard spectral density.
The diffusive behaviour for all power-law spectral densities has been catalogued in both classical and quantum cases, at both zero and non-zero temperatures  \cite{Grabert1987}, and in none of these cases does the diffusive behaviour have the shape of $\expect{\Delta Q^{2}(t)}\sim t/\log(t)$.

Subdiffusion is commonly modelled and studied in classical probability theory using continuous-time random walks (CTRWs).
In this literature it is well known that subdiffusion can be modelled by a CTRW with a jump-length distribution with a finite variance (such as a Gaussian distribution) and a heavy-tailed waiting-time distribution.
For a waiting-time distribution which scales asymptotically as $t^{-(1+\beta)}$ as $t\to \infty$ for $0 < \beta < 1$, the corresponding diffusion behaves as $\expect{\Delta Q^{2}(t)} \sim t^{\beta}$ \cite{Metzler2014}.
This is the same diffusion behaviour as seen in a generalised Langevin equation with a memory kernel with the asymptotic behaviour $k(t) \sim t^{-\beta}$ as $t\to \infty$ for $0 < \beta < 1$.

While a GLE describes a continuous process (and not piecewise constant) and therefore the full pathwise dynamics cannot be modelled by a CTRW, we might imagine a connection between the models, with a link between the memory kernel and the waiting-time distribution.
Indeed \cite{Metzler2014} presents a table of well-known models and processes which feature anomalous diffusion; however none reproduce the logarithmically corrected linear behaviour we see in our calculations.
The case of $\beta = 1$ is known to be a pathological case, with the resulting probability distribution depending on the non-asymptotic part of the waiting-time distribution \cite{Denisov2011}.
Logarithmic diffusive behaviour does occur with ultraslow processes, which can be modelled with CTRWs using a waiting time distribution of the asymptotic form $1/(t(\log t)^{1+\gamma})$ for $0 < \gamma < 1$, which result in diffusion which scales as $\expect{\Delta Q^{2}(t)} \sim (\log t)^{\gamma}$; but again this is not the logarithmically corrected linear model that we see in our damped bath model.
In fact we do not know of any classical probability model which reproduces this particular asymptotic diffusion scaling.

Some important future work will be to develop a quantum master equation or a classical Fokker-Plank equation describing the evolution of the system of interest.
While the coupling between the system and bath is identical to the Markovian Caldeira-Leggett model, this endeavour is complicated by the infinite time scale of the bath, which prohibits the usual time scale arguments that are necessary to carry out the derivation.

\section{Acknowledgements}
This work was made possible through the support of Grant 62210 from the John Templeton Foundation. The opinions expressed in this publication are those of the author(s) and do not necessarily reflect the views of the
John Templeton Foundation. For the purpose of open access, the authors have applied a Creative Commons attribution license (CC BY) to any Author Accepted Manuscript version arising from this submission.

\bibliography{references}

\end{document}